# Photonic Crystals for Nano-Light in Moiré Graphene Superlattices

S.S. Sunku[1,2], G.X. Ni[1], B.Y. Jiang[3], H. Yoo[4], A. Sternbach[1], A.S. McLeod[1], T. Stauber[5], L. Xiong[1], T. Taniguchi[6], K. Watanabe[6], P. Kim[4], M.M. Fogler[3] and D.N. Basov[1,*]

[1]Department of Physics, Columbia University, New York, NY 10027, USA

[2]Department of Applied Physics and Applied Mathematics, Columbia University, New York, NY 10027, USA

[3]Department of Physics, University of California San Diego, La Jolla, CA 92093, USA

[4]Department of Physics, Harvard University, Cambridge, MA 02138, USA

[5]Departamento de Teoría y Simulación de Materiales, Instituto de Ciencia de Materiales de Madrid, CSIC, E-28049 Madrid, Spain

[6]National Institute for Materials Science, Namiki 1-1, Tsukuba, Ibaraki 305-0044, Japan

* db3056@columbia.edu

Graphene is an atomically thin plasmonic medium that supports highly confined plasmon polaritons, or nano-light, with very low loss. Electronic properties of graphene can be drastically altered when it is laid upon another graphene layer, resulting in a moiré superlattice. The relative twist angle between the two layers is a key tuning parameter of the interlayer coupling in thus obtained twisted bilayer graphene (TBG). We studied propagation of plasmon polaritons in TBG by infrared nano-imaging. We discovered that the atomic reconstruction occurring at small twist angles transforms the TBG into a natural plasmon photonic crystal for propagating nano-light. This discovery points to a pathway towards controlling nano-light by exploiting quantum properties of graphene and other atomically layered van der Waals materials eliminating need for arduous top-down nanofabrication.

**One Sentence Summary:** Atomically relaxed twisted bilayer graphene hosts periodic arrays of topological conducting channels that act as a photonic crystal for surface plasmons.

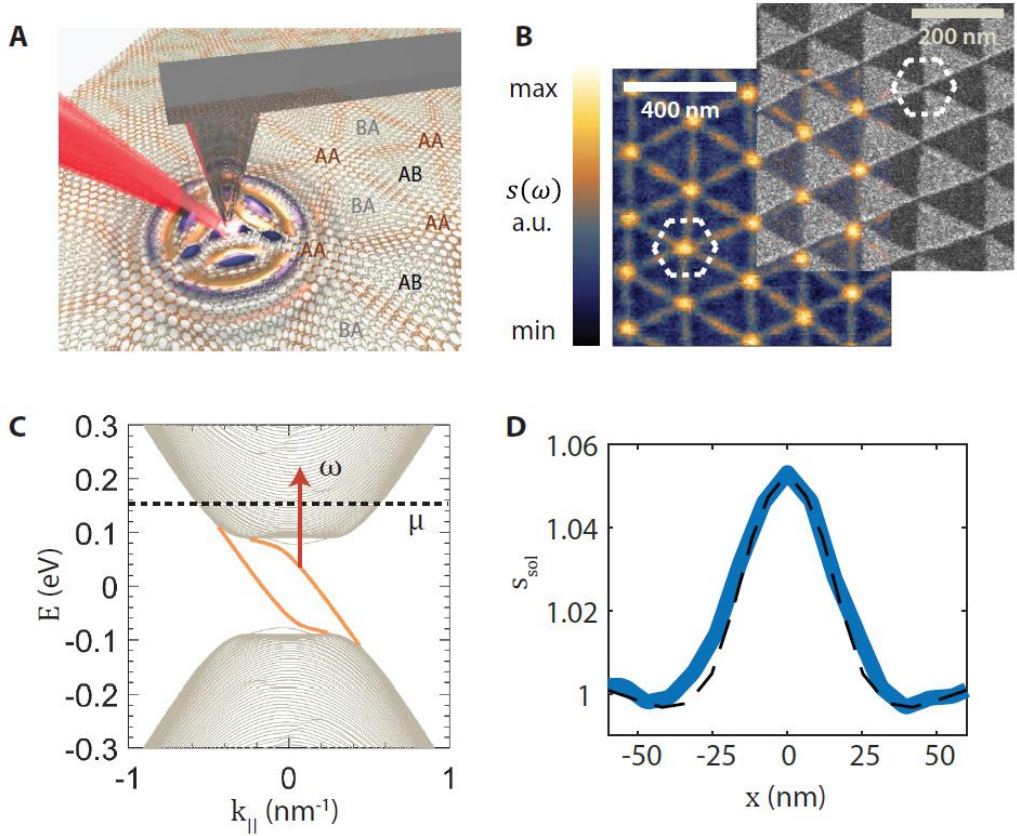

**Figure 1 | Nano-light photonic crystal formed by a network of solitons in twisted bilayer graphene.** (A) Schematic of the infrared nano-imaging experimental setup. AB, BA, AA label periodically occurring stacking types of graphene layers. (B) Left: Visualizing the nano-light photonic crystal formed by the soliton lattice. The contrast is due to enhanced local optical conductivity at solitons. Right: Dark-field transmission electron microscopy image of a twisted bilayer graphene sample showing contrast between AB and BA triangular domains. The dashed white hexagons represent unit cells of the crystals. (C) Electronic band structure of a single infinitely long soliton (only the K valley is shown). Chiral 1D states are depicted in orange. Optical transitions such as those indicated by the red arrow are responsible for the enhanced local conductivity at the location of solitons. (D) Experimental (solid) and calculated (dashed) near-field signal $s_{sol}(x)$ across a single soliton line. Calculation parameters are frequency $\omega = 1180 \text{cm}^{-1}$, Fermi energy $\mu = 0.3 \text{eV}$, interlayer bias $V_i = 0.2 \text{V}$ and dimensionless damping $\eta = 0.2$ (see text).

When light of wavelength $\lambda_0$ travels through media with periodic variations of the refractive index, one witnesses an assortment of optical phenomena categorized under the notion of a photonic crystal (*1*). The additional periodicity imposed on light can trigger the formation of a full photonic band gap (*2*) and may also produce chiral one-dimensional (1D) edge states (*3*) or exotic photonic dispersions emulating that of Dirac and Weyl quasiparticles (*4*). In principle, the photonic crystal concept is also applicable for controlling the propagation of "nano-light": coupled oscillations of photons and electrons confined to the surface of conducting media and referred to as surface plasmon polaritons (SPPs) (*5–7*). The wavelength of SPPs, $\lambda_{SPP}$, is reduced compared to $\lambda_0$ by up to three orders of magnitude (*8*). However, this virtuous confinement poses challenges for the implementation of the nano-light photonic crystals by standard top-down techniques (*9*,

*10*). Here we demonstrate a lithography-free photonic crystal for plasmons in TBG. Periodically varying optical response in these van der Waals heterostructures arises naturally due to modification of the electronic structures at moiré domain walls (solitons) formed in rotationally misaligned graphene layers (Fig. 1(B)). The most important feature of the modified electron dispersion is the appearance of chiral 1D states (one pair per valley, Fig. 1C), which are known to be topologically protected (*11*). The optical transitions involving these 1D states (vertical arrow in Fig. 1C) produce an enhancement of the local optical conductivity across the solitons (*11*). A characteristic profile of the attendant near-field signal is displayed in Fig. 1D and will be discussed below. By utilizing infrared (IR) nano-imaging experiments (Fig. 1A), we visualize the interference between SPPs propagating in solitonic networks and predict the formation of a full plasmonic band gap.

Graphene has emerged as an extremely capable plasmonic medium in view of ultra-strong confinement, quantified by $\lambda_0/\lambda_{SPP} \geq 1000$ (*8*) attained in the regime of weak loss (*12*, *13*). Plasmonic properties of graphene can be readily controlled by carrier density (*6*, *7*), dielectric environment (*12*, *14*) and ultrafast optical pulses (*15*). Here, we have explored and exploited yet another control route based on the twist angle $\theta$ between neighboring graphene layers (*14*, *16*–*20*). In TBG, the local stacking order changes smoothly across the narrow solitons separating AB- from BA- domains (*21*), as revealed (Fig. 1B) by dark field (DF) transmission electron microscopy (TEM). Previous nano-IR experiments on isolated solitons in Bernal-stacked bilayer graphene (BLG) have shown that SPPs in BLG are scattered by the solitons (*11*, *22*) analogous to the scattering of SPPs by grain boundaries in monolayer graphene (*23*). Therefore, a regular pattern of such solitons (Fig. 1(B)) is expected to act as a periodic array of scatterers thus fulfilling the key pre-condition for nano-light photonic crystal. Unlike all previous implementations of photonic crystals (*24*, *25*), our approach exploits local changes in the electronic band structure of the plasmonic medium, a quantum effect, to control optical phenomena. We explored this novel and fundamentally quantum approach for manipulating plasmons via direct nano-imaging experiments, modeling and theory.

Infrared nano-imaging (Fig. 1(A)) is central to unveiling the physics of a quantum photonic crystal for plasmons. In our experiments, infrared light at frequency $\omega = 1/\lambda_0$ is focused on the apex of a metallic tip. The amplitude of the backscattered signal $s(\omega)$ and its phase $\phi(\omega)$ are recorded using an interferometric detection (*26*). When $\omega$ is close to the optical phonon of the $SiO_2$ substrate, as in Fig. 1B, IR nano-imaging effectively reveals local variations of the optical conductivity (*26*, *27*). In Figure 1(B), we observed a six-fold pattern of bright line-like features with even stronger contrast at the intersections. A dark field TEM image of a similar TBG sample also reveals the same six-fold symmetry with features matching the nano-IR data. The periods of both patterns are consistent with the moiré length scales anticipated for a nominal twist angle of ~ 0.1°. An accurate estimate of the periodicity $a$ for our device can be directly read off the near-field image: given the observed $a \simeq 230$ nm we obtain a twist angle of $\theta \simeq 0.06°$ (*26*). We therefore conclude that the near-field image constitutes a direct visualization of the solitonic lattice.

The nano-IR contrast at the solitons is the result of topological changes to the electronic structure. When inversion symmetry is broken by an application of a perpendicular displacement field using the back gate, the Bernal stacked AB and BA domains reveal a bandgap (*28*) and the valley Chern number at K and K' valleys is ±1 (*29*). As the stacking order evolves across the soliton, the Chern numbers also change sign. The difference in Chern number leads to topologically protected one-dimensional states along the soliton (*30, 31*). The key implication of this band structure effect (*11*) is that optical transitions from the topologically protected states to empty states above the Fermi level prompt an enhanced conductivity at the soliton (Fig. 1(C)). Consistent with this view, resistivity experiments signal ballistic electron transport along the solitonic channels (*17, 32*).

Our qualitative understanding of the near-field contrast is corroborated by modeling. The near-field amplitude and phase profiles, $s_{sol}(x)$ and $\phi_{sol}(x)$, $x$ is the coordinate normal to the soliton, depend on the Fermi energy $\mu$, the interlayer bias $V_i$ and the plasmonic damping rate $\eta$ (Section S4 of (*26*)). These latter profiles obtained for isolated solitons (*11, 22*) were fully elucidated by combining electronic structure calculations, scattering theory, and numerical modeling of the tip-sample coupling (*11, 33*). Figure 1(D) shows the calculated $s_{sol}(x)$ using parameters that most closely correspond to the experiment in Fig 1(B).

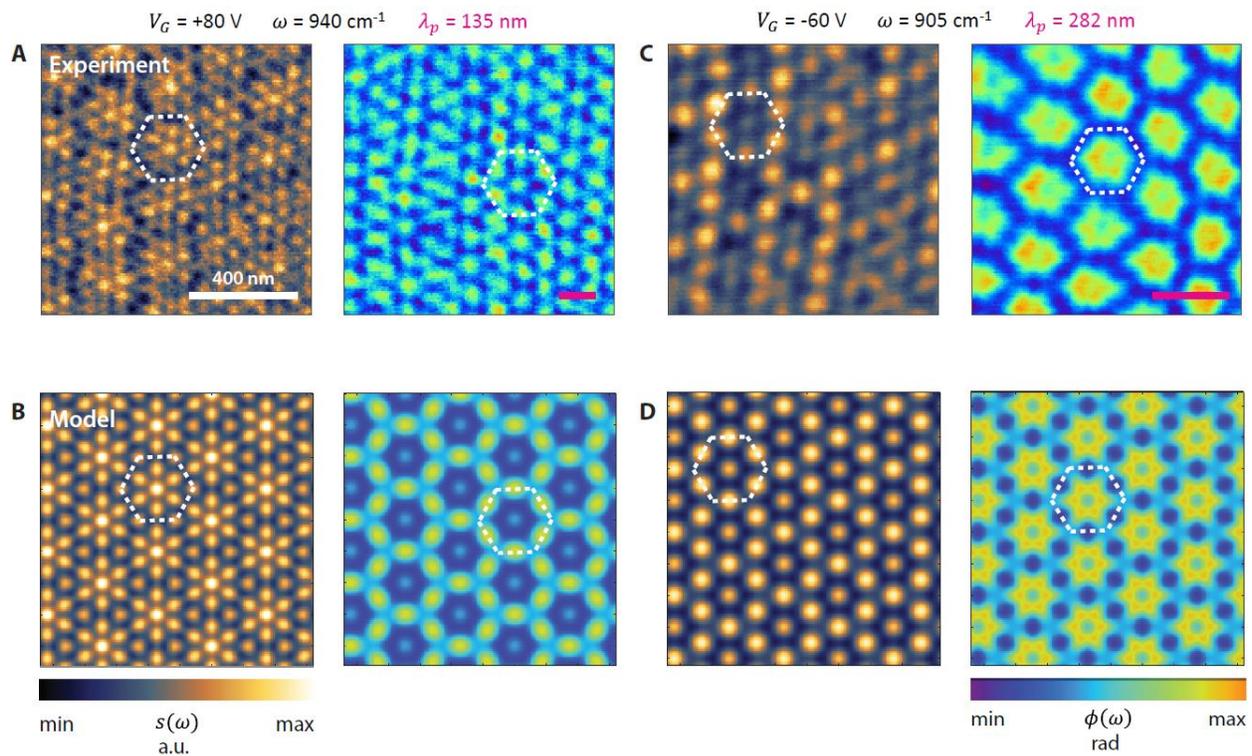

**Figure 2 | Plasmon interference patterns and superposition model analysis.** (A, C) Nano-IR images obtained for $\lambda_p = 135$ nm and 282 nm. (B, D) Near-field amplitude and phase images calculated using the superposition model (introduced in the text). The model parameters used to obtain the images are: (B) $\mu = 0.23\text{eV}, V_i = 0.3\text{V}, \eta = 0.2$ (D) $\mu = 0.35\text{eV}, V_i = 0.1\text{V}, \eta = 0.2$. The dashed-line hexagons represent the boundaries of a single unit cell and the magenta bars represent the SPP wavelengths.

We now discuss the impact of periodically varying conductivity in TBG on propagating plasmon polaritons. In our experiments, SPPs of wavelength $\lambda_p$ of the order of the soliton periodicity $a$ are introduced by the metallic tip (Figure 1(A) and Refs. (*12*, *15*)). In order to launch propagating polaritons, we choose $\omega$ to be away from phonon resonances. In this regime, the scattering of SPPs by the solitons produces fringes in both $s(\omega)$ (*6*, *7*) and $\phi(\omega)$ (*34*) corresponding to standing waves. Two-dimensional (2D) maps of both observables are displayed in Figure 2. We obtained these images in different regimes of $\lambda_p/a$ by tuning the gate voltage $V_G$ and/or $\lambda_0$. All images are dominated by maxima and minima in the nano-IR contrast, indicating the presence of constructive and destructive interference of SPPs triggered by the solitonic lattice.

The Fourier analysis of the $s(\omega)$ images shown in Figures 3 (A, B) supports our conjecture of a photonic crystal. We denote the magnitude of the 2D spatial Fourier transform of the $s(\omega)$ image as $\tilde{s}(q)$. Figure 3(A) shows $\tilde{s}(q)$ extracted from the spatially varying conductivity image displayed in Figure 1(B) and is seen to have six-fold rotational symmetry. This symmetry is preserved in the $\tilde{s}(q)$ images obtained by transforming data in Figure 2 in the regime where our structures support propagating SPPs. Figure 3(B) shows the line profiles taken along one of the high-symmetry directions for all $\tilde{s}(q)$ images. The peaks in all the images are anchored at the same momenta in Fourier space, indicating that the periodicity of the polaritonic nano-IR patterns matches that of the moiré lattice. Our nano IR imaging and its Fourier transformed patterns thus give further evidence of plasmonic interference in the soliton crystal formed in TBG.

For a quantitative analysis of the SPP interference, we introduce a superposition model. In this simplified model, we neglect multiple scattering of plasmons by these domain walls and disregard any interaction of the domain walls at their intersections. In other words, we treat the domain walls as interpenetrating and decoupled objects. We compute the near-field signal produced by a single (infinitely long) soliton as accurately as realistically possible via microscopic calculations of the electron band structure, optical conductivity, and tip-sample coupling (*11*, *35*). The superposition model takes as a basic input the profiles of the near-field amplitude $s_{sol}(x)$ and phase $\phi_{sol}(x)$ (Fig 1(D) and Section S4 of (*26*)) for a single soliton. It is easy to see that the 2D soliton lattice consists of three one-dimensional periodic arrays rotated in-plane by 120° with respect to one another. Consider one such array where solitons located at equidistant positions $x_k$. Within the superposition model, this array produces the complex near-field signal equal to the sum $\sum_k s_{sol}(x - x_k) e^{i\phi_{sol}(x - x_k)}$. Since $s_{sol}(x)$ is rapidly decreasing away from the solitons, it is sufficient to keep only a few nearest-neighbor terms in this summation. The signal from the remaining one-dimensional arrays is calculated in a similar way. The superposition of all these signals yields the images displayed in Figures 2(B) and 2(D). This procedure yields a close correspondence between the experimental data and the model in both amplitude and phase.

A key feature of moiré photonic crystal is its tunability. The periodicity of the crystal, $a$, can be continuously varied by changing the twist angle (*17*) and the SPP-soliton scattering strength can be modulated by the carrier density and the interlayer bias (*11*). In order to illustrate the tunability, we introduce the dimensionless scattering strength

$$t = \frac{1}{a} \int_{-\infty}^{\infty} dx \, \frac{\sigma_s(x) - \sigma_0}{\sigma_0}, \tag{1}$$

that governs the interaction between the SPPs and the solitons. Here $\sigma_s(x)$ is the local infrared conductivity along the direction perpendicular to the soliton and $\sigma_0$ is the asymptotic value of this conductivity far away from the soliton (*11*, *33*). Note that parameter $t$ governs the long-range behavior of the SPP waves scattered by a soliton. The details of the short-range behavior (an example of which is shown in Figs. 1(D) and 2(B), 2(D)) depend, in general, on the exact profile $\sigma_s(x)$. However, the plasmon band structure is predominantly sensitive to the long-range processes, so a single parameter $t$ suffices. We now evaluate the plasmonic band structure in momentum space for selected $t$ values using a reciprocal-space method (*26*, *36*). Figure 3(C) shows the band structure for parameters that correspond most closely to our current experiment ($a$ = 230 nm, $t = 0.02$); we notice that the plasmonic gap is insignificant. However, a larger scattering strength that is likely to be attained in future experiments does yield a full band gap arresting plasmonic propagation (Figure 3(E)). We also remark that a point-like source in plain graphene launches an isotropic cylindrical wave (Figure 3(D), left half) whose amplitude decays asymptotically as the square root of the distance. While the decay is expected to be the same for a plasmonic crystal at frequencies within the plasmonic bands, the rotational symmetry of the waves must reduce to comply with the symmetry of the crystal. The reduction to six-fold symmetry for our crystal can be revealed by dividing the signals with and without the crystal pointwise (Figure 3(d), right half). In contrast, excitations at frequencies inside the bandgap must be localized, showing exponential decay of the amplitude away from the source. We also predict that the localized states are strongly anisotropic, yielding signal distributions resembling snowflakes (Figure 3(F)) or three-pointed stars (Figure 1(A) and (*26*)). To generate patterns of this kind, one can add point-like plasmonic emitters, e.g., small gold disks (*37*) to the system.

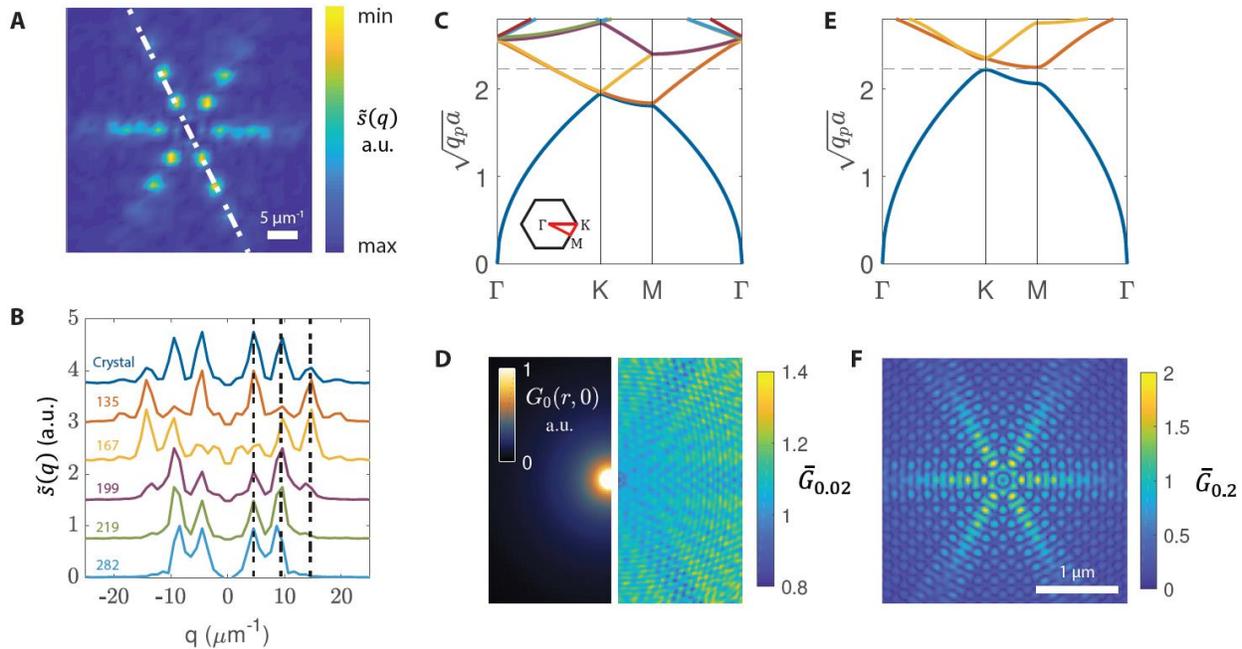

**Figure 3 | Properties of the graphene-based quantum photonic crystal.** (A) Fourier transform $\tilde{s}(q)$ of the photonic crystal image with no propagating SPPs (as in Fig. 1(B)). (B) Line profiles of $\tilde{s}(q)$ taken along the white dashed line in (A) for the crystal devoid of propagating SPPs and for the same crystal with propagating SPPs of various wavelengths $\lambda_p$. The curves are vertically offset for clarity. (C) Plasmonic band structure for dimensionless scattering strength $t = 0.02$ defined the text; $t = 0.02$ most closely corresponds to the experimentally studied crystal. (E) Plasmonic band structure for $t = 0.2$ showing the formation of a full plasmonic gap. (D) Near-field signal calculated for a point source at an AA vertex. The left half shows $G_0$, the near-field

signal computed for the empty lattice ($t = 0$). The right half depicts the ratio $\bar{G}_{0.02} = G_{0.02}/G_0$, where $G_{0.02}$ is the signal for $t = 0.02$. (F) Near-field signal ratio $\bar{G}_{0.2} = G_{0.2}/G_0$ where $G_{0.2}$ is the signal for $t = 0.2$. The frequency in both (D) and (F) corresponds to the plasmon momentum, $q_p$ that satisfies $\sqrt{q_p a} = 2.23$, shown by the dashed lines in (C, E). When this frequency is outside (inside) the band gap, the plasmonic patterns are delocalized (localized) and weakly (strongly) anisotropic, *cf.* panel D (panel F). See Section 7 of (*26*) for details of these calculations.

The nano-light photonic crystal devised, implemented and investigated in this work is unique in several ways. First, the local variation of the response is rooted in topological electronic phenomena occurring at the solitons at variance with commonplace classical photonic crystals based on locally perforated media. Second, its key parameters (periodicity and band structure) can be continuously tuned electrostatically and/or nanomechanically (*38*) and do not require extremely challenging top-down fabrication. In closing, we remark that it would be interesting to explore the regime close to the charge neutrality, where the solitons are predicted to host 1D plasmon modes (*11*, *39*). In this regime our structure would act as a 2D network or possibly, a controllable circuit capable of routing such 1D plasmons.

# Supplementary Materials for

## Photonic Crystals for Nano-Light in Moiré Graphene Superlattices

S.S. Sunku, G.X. Ni, B.Y. Jiang, H. Yoo, A. Sternbach, A.S. McLeod, T. Stauber, L. Xiong, T. Taniguchi, K. Watanabe, P. Kim, M.M. Fogler and D.N. Basov*

*Correspondence to: db3056@columbia.edu

## Materials and Methods

### Device fabrication

Twisted bilayer graphene was produced by the 'tear-and-stack' dry transfer technique as detailed in Ref *(17)*. First a layer of boron nitride (BN) is picked up using an adhesive polymer poly(bisphenol A carbonate) (PC) coated on a stamp made of transparent elastomer polydimethylsiloxane (PDMS). A large flake of monolayer graphene is identified and the BN flake is used to tear the graphene flake into two and pick up one half. The substrate is then rotated by a controlled angle and the second half of the graphene flake is picked up. The entire stack is then placed on a clean silicon dioxide/silicon substrate. The thickness of the BN used for the device in this work is 6nm.

### Infrared nano-imaging

Infrared nano-imaging was performed with a commercial scattering-type scanning near-field optical microscope (Neaspec GmbH) based on a tapping mode atomic force microscope. Our light source was a quantum cascade laser obtained from DRS Daylight Solutions, tunable from 900 cm$^{-1}$ to 1200 cm$^{-1}$. The light from the laser was focused onto a metallic tip oscillating at a tapping frequency of around 250 kHz with a tapping amplitude of around 60 nm. The scattered light was detected using a liquid nitrogen cooled HgCdTe (MCT) detector. To suppress far-field background signals, the detected signal was demodulated at a harmonic $n$ of the tapping frequency. In this work, we used $n = 4$.

## Supplementary Text

### S1. Visualizing the soliton lattice

As shown in Figure 1(B), we were able to visualize the soliton lattice by measuring $s(\omega)$ at $\omega = 1180$ cm$^{-1}$. In Figure S1(A), we show a color plot of the imaginary part of the reflection coefficient Im($r_p$) for our heterostructure when the graphene stacking configuration is Bernal stacking. By comparing the polaritonic dispersion at the three frequencies used in this work (905 cm$^{-1}$, 940 cm$^{-1}$ and 1180 cm$^{-1}$), we see that at the highest frequency, the group velocity of the polariton is highly suppressed *(27)*. Furthermore, the damping of the polariton is also highest at 1180 cm$^{-1}$. Figure S1(B) shows line profiles at 905 cm$^{-1}$ and 1180 cm$^{-1}$. The broad maximum at 1180 cm$^{-1}$ indicates that the polariton damping is high at this frequency. The combination of these two effects leads to a very short propagation length for the polariton at 1180 cm$^{-1}$ and results in the soliton appearing as a single bright line in our nano-infrared images.

The magnitude of the near-field contrast is known to be a complicated functional of the optical conductivity of the sample and the tip-sample coupling. The latter may sensitively depend on the exact geometry of the tip and other experimental parameters *(39, 40)*. While a higher local conductivity typically leads to a higher near-field signal, in general, there is no simple quantitative relation between the two. Extensive numerical modeling (similar to *(11)*) is necessary for a quantitative comparison between the experimental data and the expected $\sigma_s(x) - \sigma_0$.

Note also that "brightest" regions of the obtained images are centered at intersections of solitons (the "vertices"). Within our "superposition" approach, this property follows simply from the fact that the scattering signal at a vertex has strong contributions from all three intersecting walls. We certainly do not think that this approximation is physical, i.e., we do not think that domain walls at a vertex run straight through one another without any interaction. Nevertheless, our simulations based on the superposition approximation are in a qualitative agreement with the data, see Fig. S6. Additional dedicated experiments and theoretical calculations would be necessary to understand the structure of the vertex and its near-field response. This is a challenging problem that goes beyond the scope of the present work. Note, for example, that the plasmon wavelength is much longer than the characteristic physical dimension of the vertex. It is therefore quite difficult to resolve the microscopic structure of a vertex with the plasmon nano-imaging.

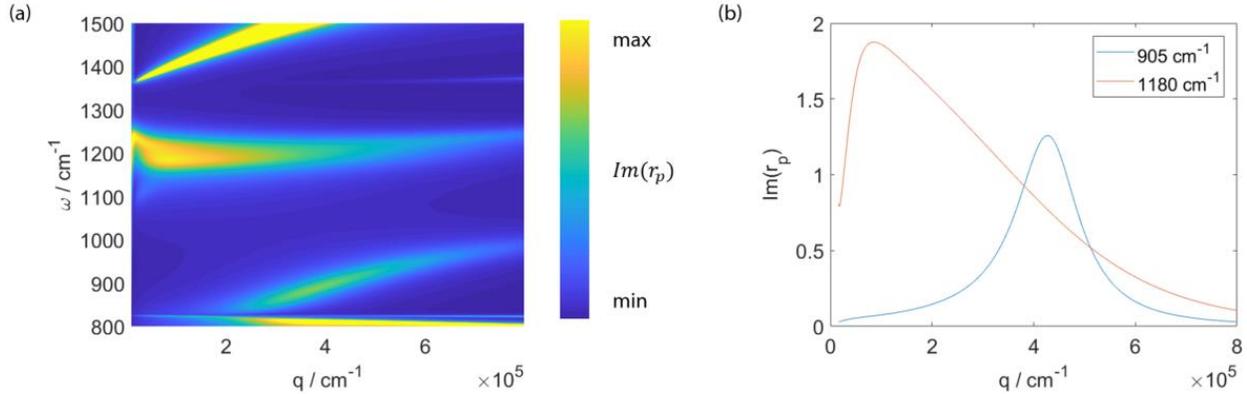

**Figure S1 | Polariton dispersion**. (a) Color plot of the imaginary part of the Fresnel reflection coefficient for p-polarized light $Im(r_p)$ for the heterostructure studied in this work when the graphene stacking is Bernal stacking. (b) Plot of $Im(r_p)$ as a function of $q$ at 905 cm$^{-1}$ and 1180 cm$^{-1}$.

## S2. Estimating the twist angle

For twisted bilayer graphene, the periodicity of the Moiré pattern $\lambda_M$ and the twist angle $\theta$ are related by $\lambda_M = a_0/[2 \sin(\theta/2)]$ where $a_0$ = 0.246 nm is the lattice constant of monolayer graphene. From Figure 1(b), we extract the periodicity to be approximately 230 nm and this corresponds to the twist angle of 0.06°. Since the Moiré periodicity is significantly longer than the inter-atomic spacing, the lattice is incommensurate despite the atomic relaxation *(41)*.

## S3. Nano-IR images at other values of $\lambda_p$

In this section, we show the nano-IR images taken at values of $\lambda_p$ other than those shown in Figure 2.

First we describe our method for extracting $\lambda_p$. We investigated a different region of the sample where we were able to images plasmons launched by a gold electrode (Figure S2). The fringes from the electrode-launched plasmons are known to have the same periodicity as the plasmon wavelength *(13, 42)*. The raw data is shown in Figure S3. At the bottom of each image, a three-peaked structure is present corresponding to a single shear soliton *(11, 22)*. The presence of this soliton to confirms that the region above the soliton is Bernal-stacked bilayer graphene with zero twist angle.

We now comment on plasmonic lifetimes and quality factors of our devices. The devices studied in this work are proof-of-concept devices. Their plasmonic lifetimes are below the record values reported in the literature *(12, 13)* but nevertheless sufficiently long for us to demonstrate rich interference patterns. To compare the lifetime in our device with values in the literature, we define the quality factor for plasmons: $Q = \text{Re}(q_p)/\text{Im}(q_p)$ where $q_p$ is the complex plasmon momentum. Based on the extracted line profiles in Figure S3, we estimate that $Q \sim 10$ in our case. This value is typical for devices where the graphene is directly on SiO$_2$ *(6, 7)*. We note that previous works on high-quality, fully hBN-encapsulated monolayer graphene (MLG) heterostructures has shown that the plasmonic lifetime is limited by the dielectric properties of the hBN and not by the graphene itself *(12, 13)*. Therefore, we expect that lifetimes similar to the record-high values reported for MLG heterostructures, $Q \sim 125$ *(13)*, can also be achieved in twisted bilayer graphene heterostructures.

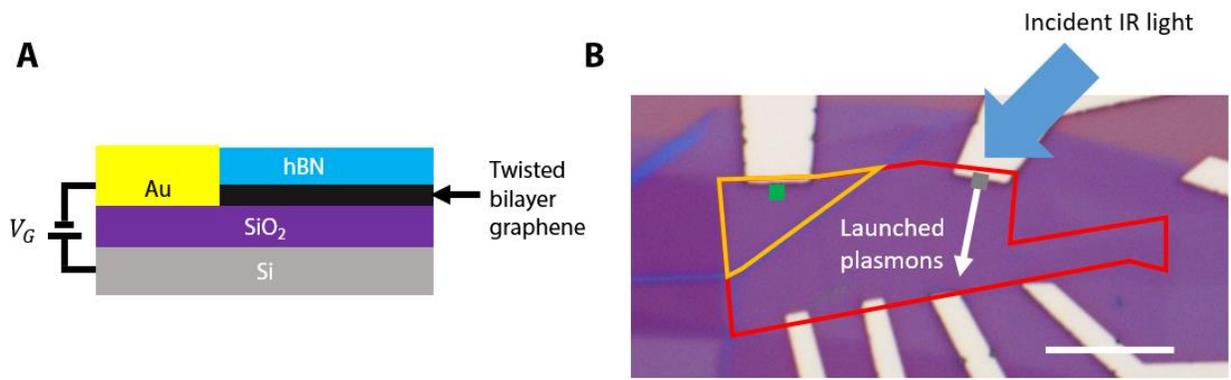

**Figure S2 | Location and direction of gold electrode launched plasmons**. (a) Schematic cross section of the device studied in this work. (b) Optical microscope image of the device. The red polygon encloses the twisted bilayer region and the orange polygon encloses the photonic crystal region. The green square indicates the region investigated in Figure 2 of the main text and the first two columns of Figure 3. The grey square indicates the region measured in the fourth column of Figure S3. Scale bar 10 µm.

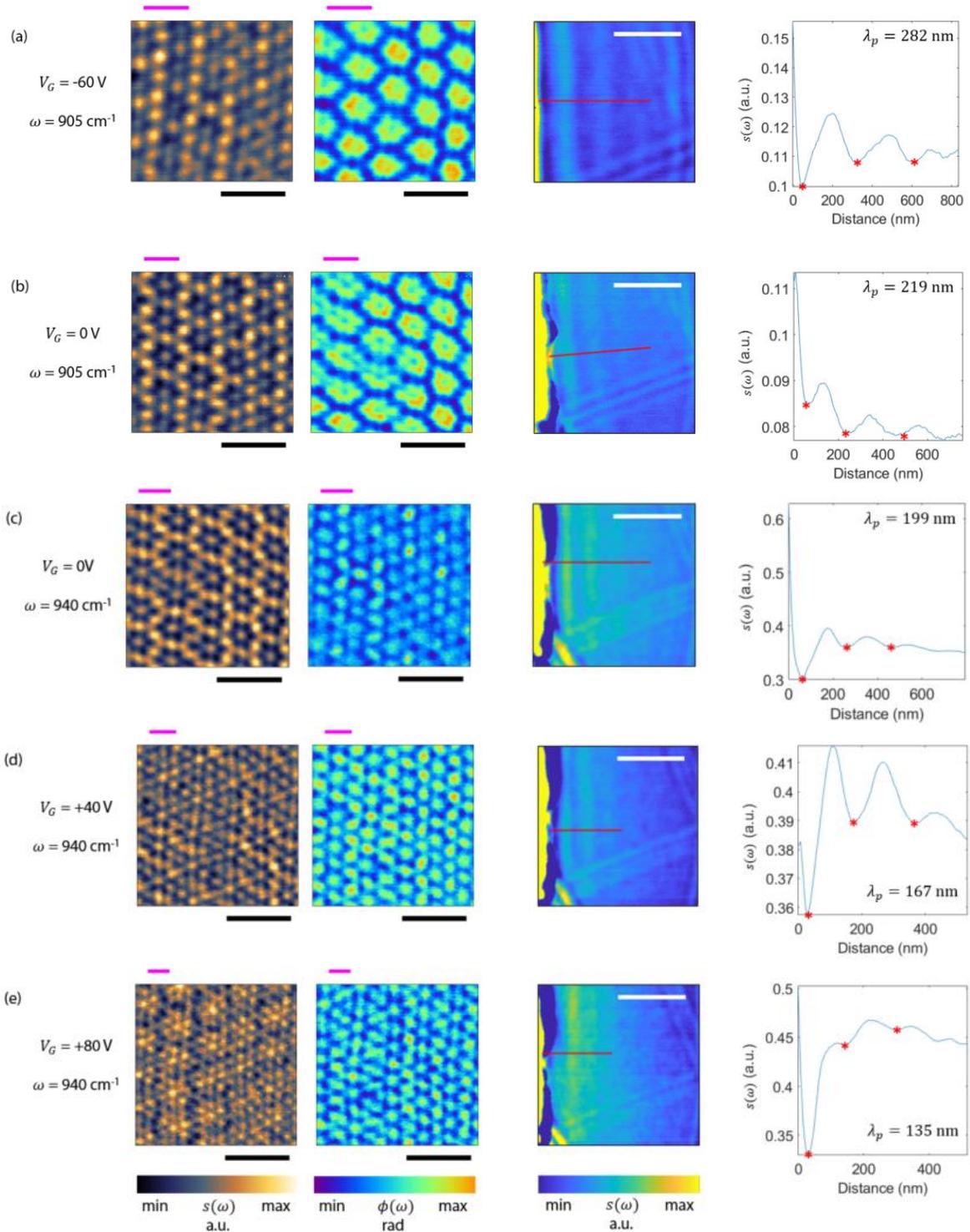

**Figure S3 | Determination of the plasmon wavelength**. The leftmost two columns show $s(\omega)$ and $\phi(\omega)$ images for various values of $\lambda_p$. Images from panels (a) and (e) are shown in Figure 2 of the main text. Black scale bar 400 nm. The magenta bars represent the SPP wavelength. The measured region is shown as a green square in Figure S2(B). The third column shows images of the electrode launched plasmons under the same conditions. The bright yellow feature visible on the left of the images in this column is the gold electrode. White scale bar 400 nm. The measured region is shown as a gray square in Figure S2(B). The rightmost column shows the line-profile taken along the red

line in the third column and averaged over a width of 80 nm. The red stars mark the first three minima. $\lambda_p$ is extracted by taking the average of the spacing between these minima.

### S4. Superposition model

The Bernal-stacked bilayer graphene regions in our sample possess an electronic gap. Inducing a bandgap in bilayer graphene requires the breaking of inversion symmetry and a single gate is sufficient for this purpose *(43-45)*. However, the limitation of a single gate is that it is not possible to independently tune the bandgap and the Fermi energy.

The model images in Figure 2 of the main text were created by superposing the simulated near-field response $s_{sol}\, e^{i\phi_{sol}}$ of a single soliton. Functions $s_{sol}(x)$ and $\phi_{sol}(x)$ were found by calculating the conductivity profile $\sigma_s(x)$ due to the soliton from the Kubo formula and using it as the input to our custom electromagnetic solver. This procedure is described in detail in our previous work *(11)* and references therein. An example of the profile $s_{sol}(x)$ we used is shown in Figure S4(C). The three adjustable parameters in this model, chemical potential $\mu$, the interlayer bias $V_i$ and the plasmonic damping rate $\eta$, were tuned until a good match with the experimental images was obtained. The scattering strength $t$ was then computed from $\sigma_s(x)$ via Eq. (1). The conductivity $\sigma_s(x)$ is in fact anisotropic and complex. Thus, instead of a single parameter $t$, one should in principle discuss two parameters, $t_x$ and $t_y$. For Figure 2(B), where $\mu = 0.23$ eV and $V_i = 0.3$ V, these scattering parameters are $t_x = -0.032 - 0.027i$ and $t_y = 0.022 - 0.015i$, which produces the band structure shown in Figure S5A. For Figure 2(d), where $\mu = 0.35$ eV and $V_i = 0.1$ V, the scattering parameters are $t_x = -0.011 - 0.011i$ and $t_y = 0.025 - 0.002i$, which produces the band structure shown in Figure S5(B). Since Figure 3 was meant to be a qualitative illustration, we presented the results for two simple isotropic examples, $t \equiv t_x = t_y = 0.02$ (panel C) and $0.2$ (panel E).

For the structures we study, it is safe to assume that the Bloch minibands are so closely spaced in energy that they do not produce any observable effects. This is because the electron Fermi wavelength is very short compared to the period of the soliton crystal: a few nanometers compared to several hundred nanometers. In other words, for electrons, unlike plasmons, the periodic potential of the soliton crystal can be considered a slowly varying, adiabatic perturbation. Therefore, the energy width of the Bloch minibands and minigaps is exponentially small. The good agreement between the experimental data and our "superposition model" further suggests that the latter model, which neglects the Bloch minibands, is nevertheless adequate in the regime of very small twist angle studied here.

The plasmon wavelength corresponding to the above Fermi energies are of the order of several nanometers. Non-local effects in the optical conductivity are only appreciable when the Fermi wavelength becomes comparable to the plasmon wavelength. Therefore, they can be neglected in our system. Such effects can become important when the plasmonic crystal is fabricated in proximity to a metallic gate *(46)*.

In principle, plasmons can be launched when incident light scatters off any inhomogeneity present in the system, e.g., the solitonic lattice itself. We do not include this in our model of the near-field contrast because it must be a negligibly small effect. Indeed, even scattering of plasmons into plasmons by the solitons is weak ($t \ll 1$). Coupling of free-space photons to plasmons must be even weaker. Indeed, a good agreement between the experiment and the model leads us to conclude that polaritons launched by the tip and/or metallic contacts dominate our images.

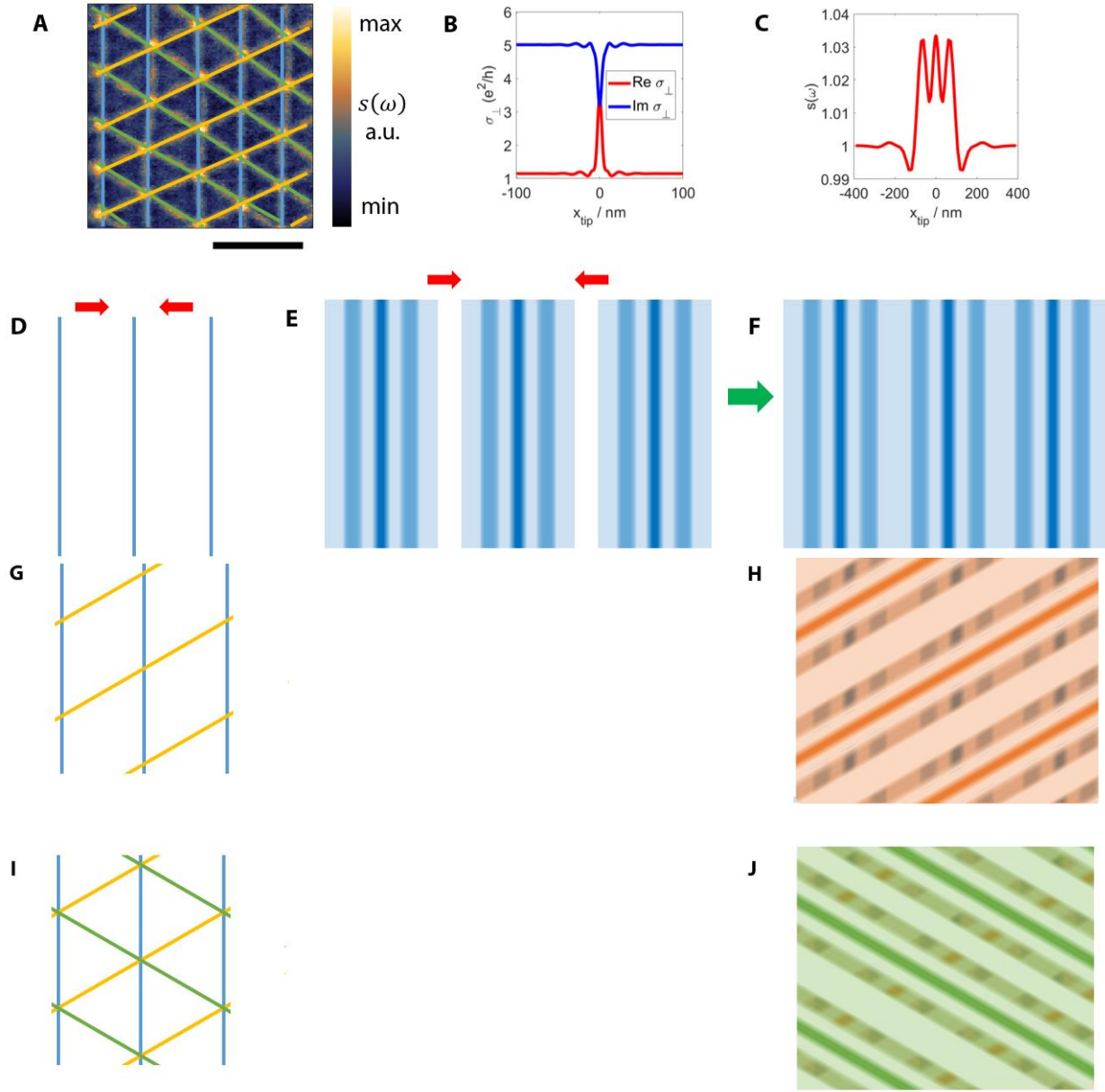

**Figure S4 | Schematic illustration of the superposition model.** (A) The soliton lattice can be decomposed into three sublattices each of which consist of solitons parallel to each other. Each sublattice is highlighted by a different set of colored lines. (B) The real and imaginary parts of the conductivity across the soliton. (C) Near-field profile due to plasmons scattering off a single isolated soliton calculated for $\mu = 0.23$eV, $V_i = 0.3$V, $\eta = 0.2$ (corresponding to Figure 2(B) in main text). (D) Schematic illustration of the solitons in the sublattice highlighted in (A). The red arrows illustrate the reduction in spacing between the solitons. (E) Schematic demonstrating the superposition of the near-field profiles as the solitons are brought closer together. (F) Schematic of the final result after superposition. (G), (H) and (I), (J) Similar schematics as (D) and (F) after the addition of the second and third sublattice respectively.

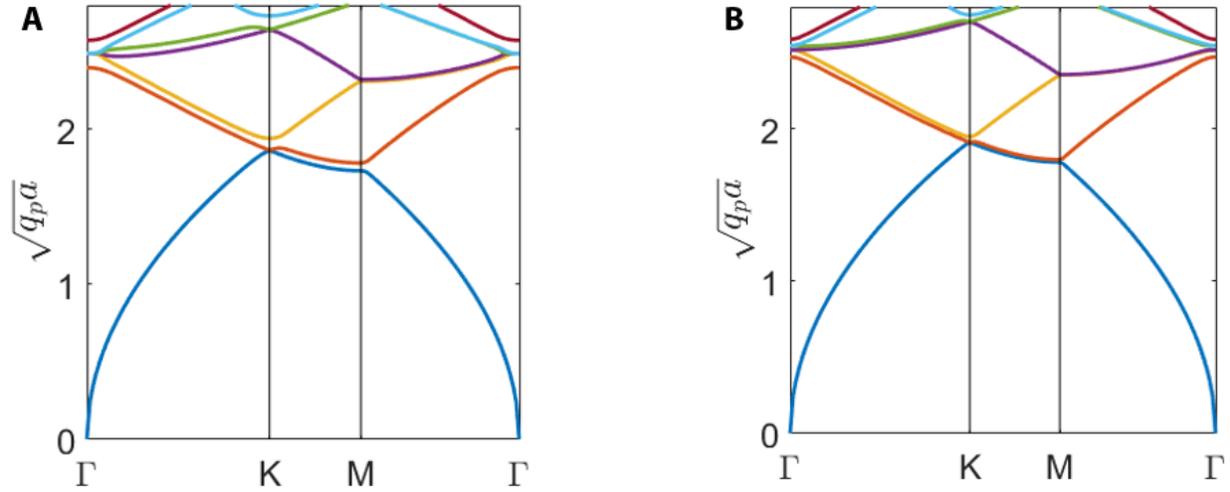

**Figure S5 | Plasmonic band structures calculated using the parameters from superposition model.** (A) Band structure corresponding to Figure 2(C) calculated using the following parameters: $\mu = 0.23$ eV, $V_i = 0.3$ V, $t_x = -0.032 - 0.027i$ and $t_y = 0.022 - 0.015i$. (B) Band structure corresponding to Figure 2(D) calculated using the following parameters: where $\mu = 0.35$ eV and $V_i = 0.1$ V, $t_x = -0.011 - 0.011i$ and $t_y = 0.025 - 0.002i$.

## S5. Comparison of line profiles from experiment and superposition model

We further compared the experimental data with the superposition model by extracting line profiles from the images. In Figure S6 (A), we label the high symmetry points of the real-space hexagonal unit cell. We then take line profiles along the paths between these points for both the experimental images and the images obtained from the superposition model. The comparison of line profiles is shown in Figure S6 (B)-(E).

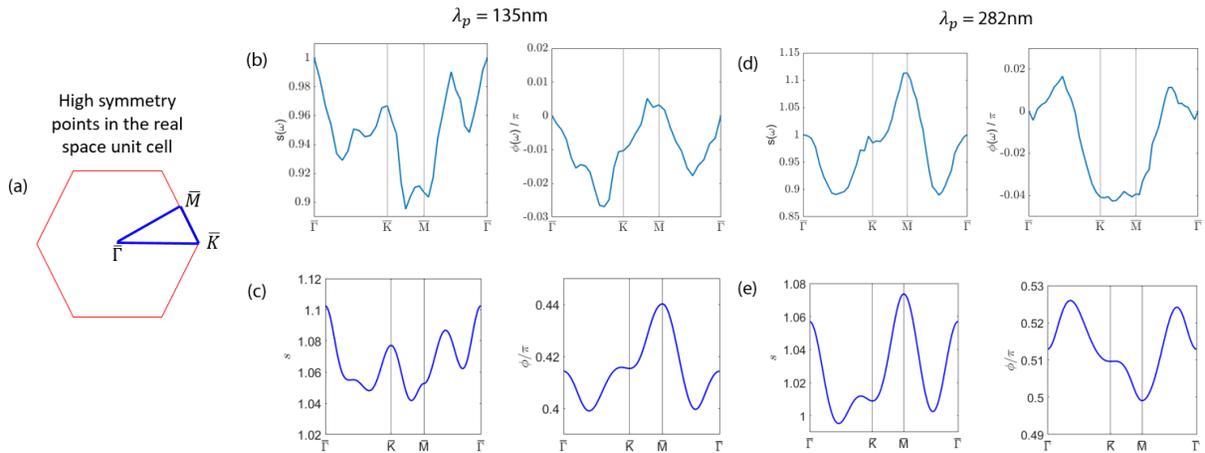

**Figure S6 | Comparison of line profiles from experiment and superposition model.** (a) Schematic showing the high-symmetry points of the real space hexagonal unit cell. (b, c) Experimental (b) and corresponding superposition model (c) line profiles for $\lambda_p = 135$nm. (d, e) Experimental (d) and corresponding superposition model (e) line profiles for $\lambda_p = 282$nm.

## S6. Fourier analysis of the scattering amplitude $s(\omega)$ images

Figures S7 (A) and S7 (B) show a comparison the Fourier transform of the near-field amplitude images obtained from experiment and the superposition model. All Fourier transforms show six-fold hexagonal symmetry, as expected from the symmetry of the crystal. Figure 4 (d) shows a plot of the line-profiles of the Fourier transforms taken along one of the high-symmetry directions. We notice that the peaks in the Fourier transforms always occur at the same positions as the peaks in the bare crystal. The only change in the line-profiles is the relative intensity of the peaks. This result shows that the periodicity of the near-field amplitude pattern at all plasmon wavelengths retains the periodicity of the bare crystal.

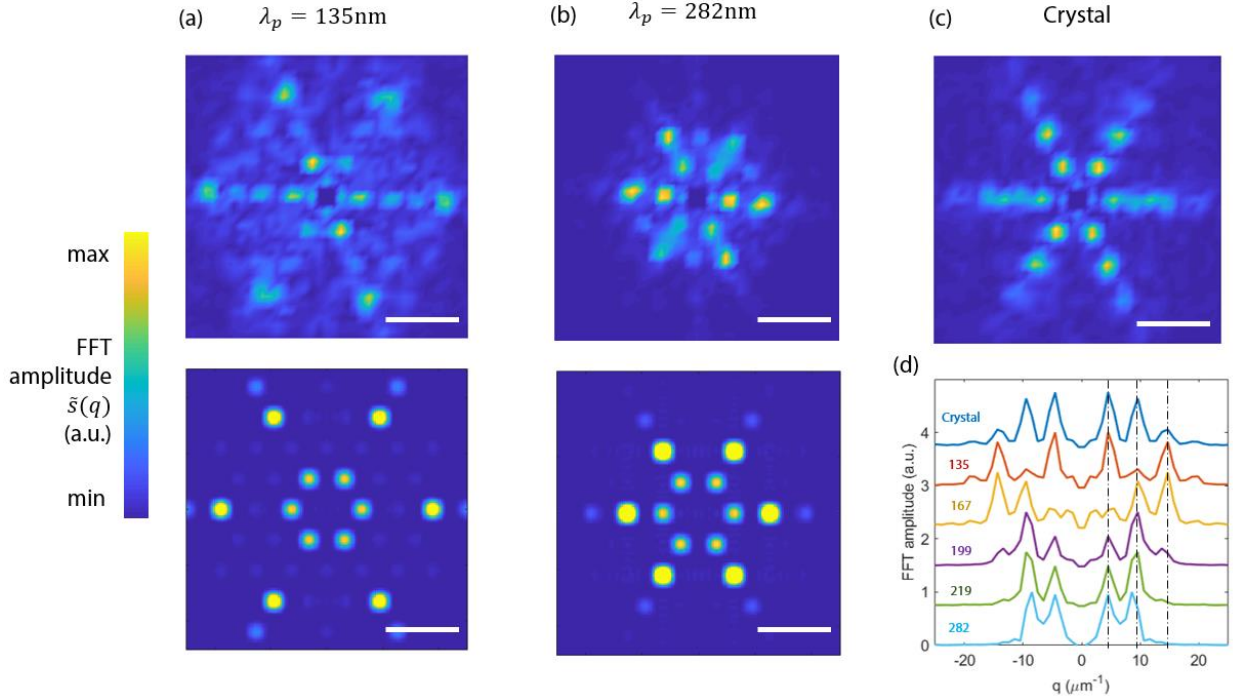

**Figure S7 | Fourier analysis of the interference patterns.** (a, b) Top: Fourier transforms of the experimental near-field amplitude, $\tilde{s}(q)$ for $\lambda_p$ of 135nm and 282nm Bottom: Fourier transforms of the near-field amplitude obtained from the superposition model. (c) Fourier transform of the bare crystal imaged experimentally (Fig 1(c)). (d) Line-profiles extracted from the experimental Fourier amplitude for the bare crystal and for different values of $\lambda_p$ taken along the white dashed line shown in (c). The black vertical dashed lines show the periodicities of the bare crystal. Scale bar 10μm$^{-1}$. The intense peak at the origin has been artificially removed from the experimental data.

## S7. Plasmonic band structure

The plasmonic band structure can be derived directly from the equation for the quasistatic plasmon potential $\phi(\vec{r})$ for a periodically varying conductivity tensor $\sigma(\vec{r})$,

$$\phi(\vec{r}) = -\left(\frac{1}{\kappa r}\right) * \vec{\nabla} \cdot \left(\frac{1}{i\omega}\vec{\sigma}(\vec{r})\vec{\nabla}\phi(\vec{r})\right), \quad \vec{r} = (x,y). \quad (S1)$$

The elements of the periodic conductivity tensor can be written in terms of reciprocal lattice vectors $Q_i$,

$$\sigma_{\alpha\beta}(\vec{r}) = \sigma_0\left(\delta_{\alpha\beta} + \sum_i \Sigma_{\alpha\beta}(\vec{Q}_i)e^{i\vec{Q}_i\cdot\vec{r}}\right), \{\alpha,\beta\} \in \{x,y\}$$

where $\sigma_0$ is the background conductivity. The potential $\phi(\vec{r})$ has the Bloch form

$$\phi_k(\vec{r}) = \sum_i c_i(\vec{k})e^{i(\vec{k}+\vec{Q}_i)\cdot\vec{r}}.$$

Equation (S1) then becomes an eigenproblem. In $k$-space,

$$HC = q_p aC,$$

where the elements of $H$ are

$$H_{ij} = \frac{1}{\sqrt{|\vec{k}+\vec{Q}_i||\vec{k}+\vec{Q}_j|}}(\vec{k}+\vec{Q}_i)\cdot[\vec{\Sigma}(Q_i - Q_j) + \delta_{ij}]\cdot(\vec{k}+\vec{Q}_j),$$

the eigenvector $C$ has elements $C_i = \sqrt{|\vec{k} + \vec{Q}_i|} \cdot c_i$, and the eigenvalue is $q_p a = \dfrac{i\kappa\omega}{2\pi\sigma_0} a$.

In our model the crystal is composed of the superposition of domain walls in three directions with a periodicity $a$, as shown in Fig. S8(a). Each domain wall is assumed to have an isotropic conductivity profile,

$$\sigma_{\text{wall}}(x_\perp) = \sigma_0 \left[1 + \frac{ta}{\sqrt{2\pi}w} \cdot \exp\left(-\frac{x_\perp^2}{2w^2}\right)\right]$$

where the wall width $w = 6$ nm (21) and $x_\perp$ is the perpendicular distance to the wall. The dimensionless scattering strength $t$, assumed constant over all frequencies, is defined as

$$t = \frac{1}{a}\int_{-\infty}^{\infty} dx \, \frac{\sigma_{\text{wall}}(x) - \sigma_0}{\sigma_0}.$$

(Same as Equation (1) of main text). We then have

$$\vec{\Sigma}(\vec{Q}_i) = t(1 + 2\delta_{Q_i,0}) \exp\left(-\frac{w^2 Q_i^2}{2}\right) \cdot I, \quad \vec{Q}_i \parallel \{\vec{P}_1, \vec{P}_2, \vec{P}_1 + \vec{P}_2\},$$

where $I$ is the identity matrix and $\vec{P}_1$ and $\vec{P}_2$ are the primitive reciprocal lattice vectors, Fig. S8(b). The eigenvalues $q_p a$ can now be calculated numerically given a grid of $\vec{Q}_i$, which is chosen large enough that the results are independent of the grid size.

The plasmon wave around a $\delta$-function impurity located at $\vec{r}'$ is given by the Green's function,

$$G(\vec{r}, \vec{r}') = \int \frac{d^2 k}{(2\pi)^2} e^{i\vec{k}\cdot(\vec{r}-\vec{r}')} \sum_i \tilde{G}_{\vec{k},\vec{k}+\vec{Q}_i} e^{-i\vec{Q}_i \cdot \vec{r}'},$$

where

$$\tilde{G}(E) = [H - (E + i0^+)]^{-1}.$$

Given an eigenvalue $E$, the location $\vec{r}'$, and the scattering strength $t$, the wave function $G(\vec{r}, \vec{r}')$ can be found. The Green's function can also be used to simulate the SNOM image, with the tip replacing the impurity as both the launcher and the detector. The SNOM signal is then $|G(\vec{r}, \vec{r})|$. To account for the finite size of the tip, we add the factor $F(\vec{k}) = k^2 e^{-2kd}$ to the integral(41) so that

$$G_{\text{tip}}(\vec{r}) = \int \frac{d^2 k}{(2\pi)^2} F(\vec{k}) \sum_i \tilde{G}_{\vec{k},\vec{k}+\vec{Q}_i} e^{-i\vec{Q}_i \cdot \vec{r}} F(\vec{k} + \vec{Q}_i).$$

Here $d = 30$ nm is the radius of curvature of the SNOM tip.

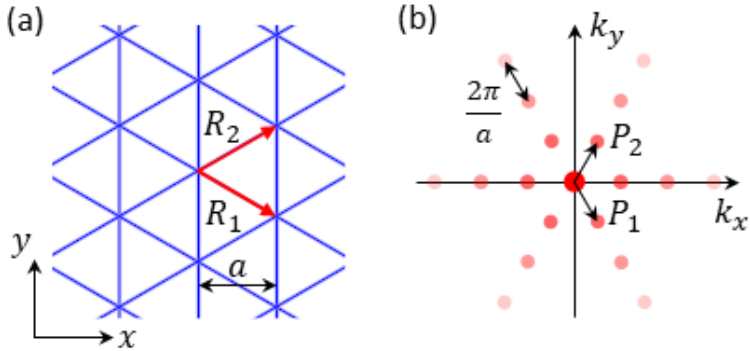

**Figure S8 | Plasmonic crystal model** (a) Lattice of domain walls in real space. $R_1$ and $R_2$ are the primitive lattice vectors, and $a$ is the periodicity of the walls. (b) In $k$-space the crystal is represented by the Bragg peaks (red dots), where $P_1$ and $P_2$ are the primitive reciprocal lattice vectors.

S8. Defect state located around the AB-region

In Figure 3(d) and 3(f) of the main text, we showed the localized modes that arise from a point defect located at the AA region. A similar source at the center of the AB region leads to a three-fold symmetric pattern, as shown in Figure S9. The difference in the patterns directly reflects the difference in the symmetry of the plasmon wavefunction around that region and are helpful in visualizing the effect of the lattice on propagating plasmons.

We further comment that the observation of the localized plasmon patterns would require improvements to the experiment. First, since $t \propto (\mu a)^{-1}$, where $\mu$ is the chemical potential of the TBG, the scattering strength can be increased by lowering $\mu$ and increasing $a$. In our experiment $\mu \simeq 0.25$ eV and $a \simeq 230$ nm, so a scattering strength of $t = 0.2$ at the same frequency can be achieved, for example, by reducing $a$ to 70 nm and $\mu$ to 0.08 eV. A top gate may also be required to maintain the same value of perpendicular displacement field. Second, plasmonic damping can be reduced by fully encapsulating the TBG and performing the experiment at low temperature *(13)*.

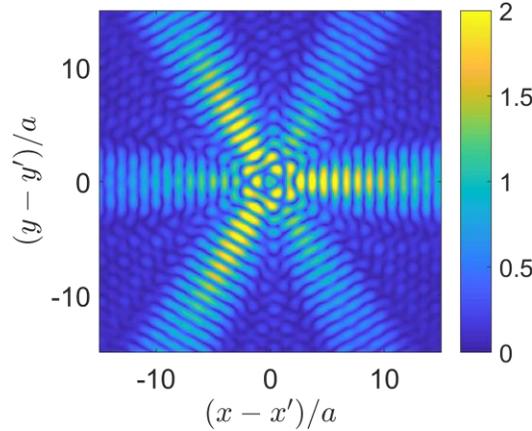

**Figure S9** | Plasmon wave function $|G(\vec{r}, \vec{r}')|$ around an impurity located at the center of the triangular AB region, $\vec{r}' = (0, 2a/3)$, normalized to the empty-lattice wave function. Parameters: $= 0.2$, $\sqrt{q_p a} = 2.23$.